\documentclass[apj]{emulateapj}

\usepackage{rotating}
\usepackage{epsfig}

\newcommand{\myemail}{rse@astro.caltech.edu}
\newcommand{\lt}{\ifmmode\,<\,\else \,$<$\,\fi}
\newcommand{\kms}{\ifmmode\,{\rm km}\,{\rm s}^{-1}\else km$\,$s$^{-1}$\fi}

\newcommand{\magarc}{\ifmmode {{{{\rm mag}~{\rm arcsec}}^{-2}}}
             \else {{{mag}$~${arcsec}$^{-2}$}}
             \fi}

\newcommand{\etal}{et~al.\@}

\newcommand{\rhosfr}{\dot{\rho}_{\star}}
\newcommand{\rhouv}{\rho_{\mathrm{UV}}}
\newcommand{\rhosfrunit}{\mathrm{M}_{\sun}~\mathrm{yr}^{-1}~\mathrm{Mpc}^{-3}}
\newcommand{\rhouvunit}{\mathrm{ergs}~\mathrm{s}^{-1}~\mathrm{Hz}^{-1}\mathrm{Mpc}^{-3}}
\newcommand{\luvunit}{\mathrm{ergs}~\mathrm{s}^{-1}~\mathrm{Hz}^{-1}}

\shorttitle{The Abundance of Star-Forming Galaxies in the Redshift Range 8.5 to 12}

\begin{document}

\title{The Abundance of Star-Forming Galaxies in the Redshift Range 8.5 to 12: New Results from the 2012 Hubble Ultra Deep Field  Campaign}


\author {Richard S Ellis\altaffilmark{1},
Ross J McLure\altaffilmark{2},
James S Dunlop\altaffilmark{2},
Brant E Robertson\altaffilmark{3},
Yoshiaki Ono\altaffilmark{4},
Matthew A Schenker\altaffilmark{1},
Anton Koekemoer\altaffilmark{5},
Rebecca A A Bowler\altaffilmark{2},
Masami Ouchi\altaffilmark{4},
Alexander B Rogers\altaffilmark{2},
Emma Curtis-Lake\altaffilmark{2},
Evan Schneider\altaffilmark{3}, 
Stephane Charlot\altaffilmark{7},
Daniel P Stark\altaffilmark{3},
Steven R Furlanetto\altaffilmark{6},
Michele Cirasuolo\altaffilmark{2,8}
}

\altaffiltext{1}{Department of Astrophysics, California Institute of 
                Technology, MS 249-17, Pasadena, CA 91125; 
                \myemail}
\altaffiltext{2}{Institute for Astronomy, University of Edinburgh, Royal Observatory,
		Edinburgh EH9 3HJ, UK}
\altaffiltext{3}{Department of Astronomy and Steward Observatory, University of Arizona, Tucson AZ 85721}
\altaffiltext{4}{Institute for Cosmic Ray Research, University of Tokyo,
                   Kashiwa City, Chiba 277-8582, Japan} 
\altaffiltext{5}{Space Telescope Science Institute, Baltimore, MD 21218}
\altaffiltext{6}{Department of Physics \& Astronomy, University of California, Los Angeles CA 90095}
\altaffiltext{7}{UPMC-CNRS, UMR7095, Institut d'Astrophysique de Paris, F-75014, Paris, France}
\altaffiltext{8}{UK Astronomy Technology Centre, Royal Observatory, Edinburgh EH9 3HJ, UK}


\begin{abstract}
We present the results of the deepest search to date for star-forming galaxies beyond a redshift 
$z\simeq$8.5 utilizing a new sequence of near-infrared Wide Field Camera 3 
images of the Hubble Ultra Deep Field. This `UDF12'  campaign completed in September 2012
doubles the earlier exposures with WFC3/IR in this field and quadruples the exposure in the key 
F105W filter used to locate such distant galaxies. Combined with additional imaging in the F140W filter, the fidelity 
of high redshift candidates is greatly improved. Using spectral energy distribution fitting techniques on objects selected
from a deep multi-band near-infrared stack we find 7 promising $z>$8.5 candidates. As none of
the previously claimed UDF candidates with $8.5<z<$10 is confirmed by our deeper multi-band
imaging, our campaign has transformed the measured abundance of galaxies in this redshift range. 
Although we recover the candidate UDFj-39546284 (previously proposed at $z$=10.3), it is undetected 
in the newly added F140W image, implying it lies at $z$=11.9 or is an intense emission line 
galaxy at $z\simeq2.4$. Although no physically-plausible model can explain the required line
intensity given the lack of Lyman $\alpha$ or broad-band UV signal, without an infrared
spectrum we cannot rule out an exotic interloper.  Regardless, our robust $z \simeq 8.5 - 10$ sample 
demonstrates a luminosity density that continues the smooth decline observed over $6 < z < 8$. 
Such continuity has important implications for models of cosmic reionization and future searches 
for $z>$10 galaxies with JWST.
\end{abstract}


\section{Introduction}\label{sec:intro}

Good progress has been achieved in exploring the latest frontier in cosmic
history, namely the 700 Myr period corresponding to the redshift interval $6<z<15$. During this time, 
star-forming galaxies likely played a significant role in completing the reionization of intergalactic hydrogen 
\citep{robertson2010a,bromm2011a,dunlop2012a}. Inevitably, our census of galaxies during this era is limited by 
our current observational facilities. Most progress has been made in the lower redshift range $6<z<8.5$ 
via deep imaging with the {\it Hubble Space Telescope} (HST). This has revealed several hundred star-forming galaxies 
and a dominant contribution to the luminosity density from low luminosity examples \citep{oesch2010a,mclure2011a, 
bouwens2010a, bouwens2012a}. Measures of the assembled stellar mass from {\it Spitzer Space Telescope}
photometry at $z\simeq$5-6 \citep{stark2007a,eyles2007a,gonzalez2010a,gonzalez2011a,labbe2012} suggest
that star formation extended to redshifts well beyond $z\simeq$8 but there has been limited progress in 
finding these earlier, more distant, sources.

Various groups have attempted to find $z>8.5$ galaxies using the well-established technique of 
absorption by intervening neutral hydrogen below the wavelength of Lyman $\alpha$. 
A redshift $z$=8.5 represents a natural frontier corresponding to sources which progressively  `drop out' in the 
HST $Y$-band F105W and $J$-band F125W filters. \citet{bouwens2011a} and \citet{yan2010a} used data from the 
campaign completed in 2009 with the near-infrared Wide Field Camera 3 (WFC3/IR) in the Hubble Ultra 
Deep Field (GO 11563, PI: Illingworth, hereafter UDF09). Bouwens \etal\ initially located 3 promising $J$-band dropouts
at $z\simeq$10 but, on re-examining the completed dataset, presented only a single candidate, 
UDFj-39546284 at a photometric redshift of $z$=10.3, not drawn from the original three. 
Bouwens \etal\ also found 3 sources in the interval $8.5<z<9$, robustly detected in F125W with (F105W - F125W) 
colors implying a Lyman break near the red edge of the F105W filter.  In marked contrast, Yan \etal\ presented a list of 20 
faint $J$-band dropout candidates  arguing all had redshifts $z>8.5$. However, none of the Yan \etal\ and 
Bouwens \etal\ candidates are in common.  Gravitational lensing by foreground clusters of galaxies can overcome some of the
difficulties associated with deep imaging of blank fields. Such sources can be
magnified by factors of $\times$5-30 ensuring more reliable photometry \citep{richard2011a}. In 
favorable cases, their multiply-imaged nature offers a lower limit on their angular diameter 
distance \citep{ellis2001a, kneib2004a}. The CLASH HST survey  (GO 12065 - 12791, PI: Postman) has 
discovered several such $z>8.5$ candidates, three at $z\simeq$9-10\citep{zheng2012a, bouwens2012b} 
and a multiply-imaged source at $z$=10.7\citep{coe2012a}.

A key issue is the uncertainty in converting the various detections into estimates of the 
abundance of galaxies beyond $z\simeq$8. \citet[][see also \citet{oesch2012a}]{bouwens2011a}
claimed that their detection of a single $z\simeq$10.3 candidate in the UDF09 campaign implies a
shortfall of a factor $\simeq$3-6 compared to that expected from the declining star formation rate
density over $6<z<8$. This could imply the growth of activity was particularly rapid during the 
200 Myr from $z\simeq$10 to 8. However, \citet{coe2012a} claim the CLASH detections are 
consistent with a continuous decline to $z\simeq$10.7. One limitation of the lensing strategy 
as a means of conducting a census (rather than providing individual magnified sources for 
scrutiny) is the uncertainty associated with estimating the survey volume which depends 
sensitively on the variation of magnification with position across the cluster field \citep[c.f.,][]{santos2004a,stark2007b}.

There are several drawbacks with the earlier UDF09 campaign with respect to conducting
a census of $z>$8.5 galaxies. Limiting factors in considering the robustness of the faint 
candidates include the poor signal to noise in subsets of the F160W data, 
the reliance on only a single detection filter and the limited depth of the critical 
F105W imaging data whose null detection is central to locating $z>8.5$ candidates. 

This article heralds a series that presents results from a deeper UDF campaign
with WFC3/IR completed in September 2012 (GO 12498, PI; Ellis, hereafter UDF12)
which remedies the above deficiencies by (i) substantially
increasing the depth of the F105W image (by $\times$4 in exposure time) essential for 
robust rejection of $z<$8.5 sources, (ii) increasing the depth of the detection filter F160W 
(a 50\% increase in exposure time) and (iii) adding a deep image in the F140W filter
matching the depth now attained in F160W. This filter partially straddles the F125W and F160W 
passbands offering valuable information on all $z>7$ sources, the opportunity for an 
independent detection for $8.5<z<10.5$ sources and the first dropout search 
beyond $z\simeq$10.5.

The UDF12 survey depths (including UDF09) in the various filters are summarized in Table 1. 
Our aim, achieved in full, has been to match the depths in 
F125W, F140W, and F160W for unbiased high redshift galaxy detection, and to reach 
0.5\,mag deeper in F105W to ensure a 2-$\sigma$ limit 1.5\,mag deeper than the 
5-$\sigma$ limit in the longer wavelength bands. 
Further details of the survey and its data reduction are provided in \citet{koekemoer2012a} and catalogs
of $z\simeq$7 and 8 sources used to estimate the luminosity function are presented
in complementary articles by \citet{schenker2012a} and \citet{mclure2012a}. 
The spectral properties of the high-redshift UDF12 sources are measured and analyzed by
\citet{dunlop2012b}. A review
of the overall implications of the survey in the context of cosmic reionization is provided in
\citet{robertson2012a}. Public versions of the final reduced WFC3/IR UDF12 images, 
incorporating additions of all earlier UDF data, are available to the community on the team web 
page\footnote{\rm http://udf12.arizona.edu/}. All magnitudes are in the 
AB system \citep{oke1974}.

\section{Star Forming Galaxies with $z>$8.5}\label{sec:candidates}

To select $z>$8.5 candidates, we examined the stacked combination of the 80 orbit F160W (UDF12 plus UDF09), 
30 orbit F140W (UDF12) and 34 orbit F125W (UDF09) images and located all sources to a 5$\sigma$ limit 
within filter-matched apertures of $0.4-0.5$ arc sec corresponding to $m_{AB}\simeq$29.9-30.1. 
Making effective use of our new ultra-deep 93 orbit (71 from UDF12, 22 from UDF09) F105W image and
the deep ACS photometry, we utilized the SED approach discussed in \citet{mclure2010a,mclure2011a} to
derive photometric redshifts of all such sources. Seven convincing $z>8.5$ candidates were found. 
An independent search using the same master sample selecting those which drop out in F105W (2$\sigma$ rejection 
corresponding to $m_{AB}>31.0$) and no detection (2$\sigma$) in a combined ACS $BViz$ stack 
delivered the same $z>$8.5 candidates. All sources but one (see below) are detected in more 
than one filter and all are detected with an appropriately-reduced signal/noise in time-split subsets over the 
collective UDF09 and UDF12 campaigns.  Figure 1 shows HST broad-band images for these 7 sources. Their 
SED fits and redshift probability distributions $p(z)$ are given in Figure 2. Identifications, source photometry 
and optimum redshifts are summarized in Table 1. 

The great advantage of the SED fitting approach is that it allows us to quantify the possibility of alternative
low-redshift solutions. Four of our 7 objects (UDF12-3921-6322, UDF12-4265-7049, UDF12-4344-6547 \& UDF12-3947-8076) 
have low probabilities of being at $z<4$ ($1-4\%$). UDF12-4106-7304 has a $\simeq10\%$ probability for $z<4$ and lies close to the 
diffraction pattern of an adjacent source which may affect the F140W photometry (Figure 1). UDF12-3895-7114 
is the least secure with a 28\% probability of lying at $z<4$. We discuss UDF-3954-6284 below.

Our deeper F105W data and the new F140W image also enables us to clarify the nature of
$z>8.5$ sources claimed in the earlier UDF09 analyses (see Table 1). In \citet{mclure2011a}'s UDF09 analysis, 
no robust $J$-band dropout source was claimed  \citep[see also][]{bunker2010a}. 
However a solution with $z$=8.49 was found for HUDF\_2003 which was also listed as 
the brightest extreme $Y$-band dropout in \citet{bouwens2011a}  (ID: UDFy-38135539) who inferred a
redshift $z\approx8.7$\footnote{This source was examined spectroscopically using the VLT SINFONI integral field spectrograph
by \citet{lehnert2010a} who reported a detection of Ly$\alpha$ emission at $z$=8.6 but this 
claim is refuted by Bunker et al (in preparation) following a separate spectroscopic 
exposure with the higher resolution spectrograph X-shooter.} Our new SED analysis indicates this source 
is at $z$=8.3. Similarly, two further extreme $Y$-band 
dropouts listed by \citet{bouwens2011a} - UDFy-37796000 and UDFy-33436598 at 
redshifts of $z\approx8.5$ and 8.6 now lie at $z$=8.0 and 7.9, respectively. 
\citet{bouwens2011a} initially presented 3 sources as promising $J$-band dropouts
(see Table 1). Two of these are detected in our deeper F105W data and lie at 
lower redshifts (UDFj-436964407 at $z$=7.6 and UDFj-35427336 at $z$=7.9,
although $z\simeq$2 solutions are also possible). One $Y$-band dropout in \cite{bouwens2011a},
UDFy-39468075 moves into our sample at $z$=8.6. Finally, UDFj-38116243 claimed in the 
first year UDF09 data but later withdrawn by \citet{bouwens2011a} is below our 5 $\sigma$ detection limit.
\citet{yan2010a} listed 20 potential $J$-band dropout candidates.  Inspection of
these revealed no convincing $z>8.5$ candidates; most appear as tails of bright objects 
and cannot be reliably photometered by SeXtractor. All of the `Y dropouts' claimed by 
\cite{lorenzoni2011} have robust F105W detections in our deeper data and lie below $z$=8.5. 

In summary, only one object claimed to be at $z>8.5$ from the earlier UDF09 analysis
remains and that is the final $J$-band dropout presented by \citet{bouwens2011a} at 
$z$=10.3, UDFj-39546284 ($\equiv$UDF12-3954-6384 in Table 1). However, its non-detection 
in the UDF12 F140W data indicates a yet higher redshift of $z$=11.9 (Figure 2). The most significant
advance of our campaign is a significant increase (from 0 to 6) in the number of robustly determined 
UDF sources in the redshift range $8.5<z<$10. 

\subsection{Contamination from Strong Emission Line Sources?}\label{sec: emission lines}

A major motivation for the additional F140W filter in our UDF12 strategy was
to ensure the robust detection in two filters of potential $8.5<z<11.5$ candidates 
since the flux above 1216 \AA\ would be visible in both filters. This is the case for all 
but one of our UDF12 candidates (Table 1). A major surprise is the non-detection in F140W of 
UDFj-30546284 implying a redshift of $z$=11.90 (Figures 1 and 2).

Single band detections are naturally less convincing, although UDFj-30546284
is confirmed in F160W sub-exposures through UDF09 and UDF12, leaving no doubt
it is a genuine source. However, an alternative solution must also be carefully considered. 
The sharp drop implied by the F140W - F160W $>$1.5 (2$\sigma$) color precludes any reasonable
foreground continuum source (Figure 2) but a possible explanation might be the presence of a 
very strong emission line. Recent WFC3/IR imaging and grism spectroscopy of $z\simeq$2 
galaxies has revealed a population of extreme emission line galaxies (EELGs).  
\citet{van_der_wel2011a} have identified an abundant population of  EELGs at $z\simeq$1.7 in 
the CANDELS survey using purely photometric selection techniques. Spectroscopy 
of a subset has verified the presence of sources with rest-frame [O III] equivalent widths 
up to $\simeq$ 1000 \AA\ . Independently, \citet{atek2011a} located a similar population in the 
WISP survey over $0.35<z<2.3$ and comment specifically that such sources could contaminate
dropout searches.

Following techniques described in \citet{robertson2010a} and \citet{ono2010a},
we have simulated model spectra for young low metallicity dust-free galaxies including the 
contribution from strong nebular lines. Figure 3a shows the expected F105W - F160W color
as a function of redshift for starbursts with ages of 1 and 10 Myr demonstrating that it 
is not possible to account for the significant excess flux in F160W from either intense 
[O II] 3727 \AA\ at $z\simeq$3.4 or [O III] 5007 \AA\ at $z\simeq$2.4. Figure 3b illustrates, for the 
case of intense [O III]  emission that the expected stellar plus nebular emission spectrum of 
a 10 Myr starburst would violate the photometric flux limits provided by the various 
ACS and WFC3/IR broad band non-detections. If all of the emission in F160W arises from [O III]
above a blue $\beta$=-2 stellar continuum, the rest-frame equivalent width would have to be 
$>$4500 \AA\ , beyond that of any known object. A further difficulty is the absence of the expected 
Lyman $\alpha$ emission following a recent Keck optical spectrum (Figure 3b). 
However, although we can find no physically self-consistent starburst model that can simultaneously 
explain the F160W emission and satisfy our upper limits in Figure 3b, only an infrared spectrum 
would completely eliminate the possibility of some exotic foreground emission line source. 
As the source has $H_{AB}$=29.3, this would be a very challenging observation. For
the remainder of the paper we will interpret this source with caution.

\section{The Abundance of Galaxies with $8.5<z<12$}\label{sec:abundance}

A key issue is whether the declining cosmic star formation which is now well-established 
over $6<z<8$ \citep{bouwens2007a} continues to higher redshift as suggested by the 
presence of evolved stellar populations with ages of $\simeq$200-300 Myr at 
$z\simeq$5-7 \citep[e.g.,][]{richard2011a}. \citet{bouwens2011a} claimed, from their 
detection of apparently only one object at $z\simeq$10 c.f. three expected, that 
the star formation history declines more steeply beyond $z\simeq$8 \citep[see also][]{oesch2012a} 
to $\rhosfr(z\sim10) \approx 2\times10^{-4}~\rhosfrunit$. Recently, the CLASH survey 
has located several $z>$8.5 candidates\citep{coe2012a} implying star formation rate densities 
higher than claimed by \citet{bouwens2011a}  at $z\sim10$. However, the uncertain search volumes 
inherent in the lensing method are a major concern.

In Figure \ref{fig:rho_uv} we present the implications of the significant increase in the
number of $8.5<z<12$ sources arising from the UDF12 campaign. Our SED-based 
selection method enables us to consider separately four redshift bins.
As a direct determination of the luminosity function at $z>8.5$ is not yet possible, to 
estimate the UV luminosity densities for our four detections at $8.5\lesssim z<9.5$
we calculate the required redshift  evolution in the characteristic luminosity $dM_\star/dz$ 
such that a survey of our depth and selection efficiency would recover the number of sources 
we find. This calculation is performed assuming simple luminosity evolution from $z\sim8$, 
keeping the luminosity function normalization $\phi_{\star}$ and faint-end slope $\alpha$ fixed 
at the $z\sim8$ values measured by \citet{bradley2012a}. To reproduce our sample with mean redshift
$\langle z\rangle\approx8.9$, we find that $dM_{\star}/dz\approx1.09$.
The luminosity density can then be estimated by integrating the projected luminosity function 
parameters to $M_{\mathrm{UV}}\approx-17.7$AB \citep[e.g.,][]{bouwens2011a,coe2012a,bouwens2012b}.  We find
 $\rhouv(z\sim8.9)\approx1.08\times10^{25}~\rhouvunit$ (Figure \ref{fig:rho_uv}, blue point).  
A similar calculation provides $\rhouv(z\sim9.8)\approx7.89\times10^{24}~\rhouvunit$
from the two $z\sim9.5$ detections (magenta point). The expected UDF cosmic variance 
for $8.5\lesssim z\lesssim9.5$ is $>$40\% \citep{robertson2010b}. Within $10.5\lesssim z\lesssim11.5$, 
we find no candidates.  Nonetheless we can use the same methodology to provide an upper limit 
of $\rhouv(z\sim10.8)<9.57 \times 10^{24}~\rhouvunit$ (Figure \ref{fig:rho_uv}, purple upper limit).  

Considering the putative $z\sim12$ source, both its morphology (Figure 1) and its luminosity
cause us to be cautious, particularly given the paucity of other detections beyond $z\simeq$10.5.
Nonetheless, since the emission line hypothesis is equally difficult to accept (Section 2.1), 
we estimated the luminosity density using only the source luminosity ($M_{UV}=-19.6$AB accounting 
for IGM absorption in F160W, or $\log_{10} L_{\mathrm{UV}} = 28.48~\log_{10}~\luvunit$ ) 
and the UDF survey volume $V(11.5\lesssim z\lesssim12.5)=6.37\times10^{3}~\mathrm{Mpc}^{3}$.
The resulting luminosity density $\rhouv(z\sim11.8)>4.7\times10^{24}~\rhouvunit$ is thus
a lower limit (Figure \ref{fig:rho_uv}, red point), and conservatively does not include multiplicative effects
of selection efficiency or involve extrapolations from the $z\sim8$ luminosity function.
An additional possibility is that the F160W is contaminated by Ly$\alpha$ emission. The additional $z$=12
point (yellow) illustrates how this limit would be affected for a rest-frame equivalent width of 260 \AA\
of which half is absorbed by neutral hydrogen.

In summary, the new galaxy sample provided by UDF12 has enabled us to present the first 
meaningful estimate of $\rho_{UV}(z)$ beyond $z \simeq 8.5$. The six galaxies with $8.5 < z < 10$ 
indicate a modest shortfall in $\rho_{UV}(z)$ beyond a simple extrapolation of the
trend at $6 < z < 8$ (less sharp than that suggested by \citet{bouwens2011a}, but below 
(albeit consistent with) the cluster results \citep{zheng2012a,coe2012a}). However, 
if UDFj-30546284 is genuinely a $z$=12 galaxy (and does not have substantial Lyman-$\alpha$ emission) 
then we have witnessed an even more measured decline in $\rho_{UV}(z)$ to the highest redshift yet probed.

\section{Discussion}\label{sec:discussion}

The UDF12 data has demonstrated the continued effectiveness of HST to undertake
a census of very high redshift galaxies. Our discovery of the first robust sample of galaxies with
$z>8.5$ and possibly the most distant galaxy at $z\sim12$ extends HST's reach
further into the reionization epoch than previously thought possible \citep[c.f.,][]{bouwens2011a}.  
While the question of whether star-forming galaxies were solely responsible for reionizing intergalactic 
hydrogen is more reliably addressed through precise constraints on the $z\sim7-8$ luminosity 
function faint end slope \citep[][see \citealt{robertson2012a}]{schenker2012a,mclure2012a}
this work has placed the first constraint on the SFR density only 360 million years after
the Big Bang. Evidence for actively star-forming galaxies significantly beyond the instantaneous 
reionization redshift $z_{\mathrm{reion}}\approx10.6\pm1.2$ implied by observations of the cosmic
microwave background \citep{komatsu2011a} motivates future observations with {\it James Webb 
Space Telescope}. Our estimate of the $z\sim10$ star formation rate densities are 
consistent with previous analyses aimed at explaining the measured Thomson optical depth 
\citep[e.g.][]{robertson2010a,kuhlen2012a} and that required to produce the stellar masses of $z<8$
sources observed by {\it Spitzer} \citep[e.g.,][]{stark2012a,labbe2012}.  Our results remain consistent 
with the simple picture for the evolving star formation rate density,  stellar mass density, Thomson optical depth, 
and IGM ionization fraction presented in \citet{robertson2010a}.

\acknowledgements

US authors acknowledge financial support from the Space Telescope Science Institute under award HST-GO-12498.01-A. 
JSD acknowledges support of the European Research Council and the Royal Society. RJM acknowledges funding
from the Leverhulme Trust. We thank Kimihiko Nakajima for assistance with the Keck spectroscopy.
This work is based on data from the {\it Hubble Space Telescope} 
operated by NASA through the Space Telescope Science Institute via the association of
Universities for Research in Astronomy, Inc.\@ under contract NAS5-26555.

\begin{deluxetable*}{lcccccccl}
\tablecolumns{9}
\tablewidth{0pt}
\tablecaption{\bf $z>8.5$ Candidates\label{tab:z_8_5_candidates}}
\tablehead{
\colhead{ID} & \colhead{RA} & \colhead{Dec}
& \colhead{$z_{\rm SED} (\pm 1\sigma)$} & \colhead{$Y_{105W}$} & \colhead{$J_{125W}$} & \colhead{$J_{140W}$} & \colhead{$H_{160W}$} & \colhead{Notes}}
\medskip
\startdata
\multicolumn{9}{c}{\bf UDF12 Survey Depth 5-{\bf $ \sigma$} AB (aperture diameter arcsec - 70\% enclosed point source flux)}\medskip \\
& & & & 30.0 (0.40) & 29.5 (0.44) & 29.5 (0.47) & 29.5 (0.50) &  \medskip\\

\multicolumn{9}{c}{\bf UDF12 Galaxies\tablenotemark{a}} \smallskip \\
UDF12-3954-6284 & 3:32:39.54 & -27:46:28.4 & 11.9 $^{+0.3}_{-0.5}$ & $>$ 31.2 & $>$ 30.7 & $>$ 30.5 & 29.3 $\pm$ 0.2 & UDFj-39546284 B11\tablenotemark{b} \\
UDF12-4106-7304 & 3:32:41.06 & -27:47:30.4 & 9.5 $^{+0.4}_{-0.8}$ & $>$ 30.8 & $>$ 30.0 & 29.8 $\pm$ 0.3 & 29.7 $\pm$ 0.3 & \\
UDF12-4265-7049 & 3:32:42.65 & -27:47:04.9 & 9.5 $^{+0.4}_{-0.7}$ & $>$ 31.2 & 30.4 $\pm$ 0.6 & 29.9 $\pm$ 0.4 & 29.7 $\pm$ 0.4 & \\
UDF12-3921-6322 & 3:32:39.21 & -27:46:32.2 & 8.8 $^{+0.4}_{-0.2}$ & $>$ 31.2 & 29.9 $\pm$ 0.3 & 29.6 $\pm$ 0.3 & 29.9 $\pm$ 0.3 & \\
UDF12-4344-6547 & 3:32:43.44 & -27:46:54.7 & 8.8 $^{+0.5}_{-0.5}$ & $>$ 31.2 & 30.0 $\pm$ 0.3 & 30.1 $\pm$ 0.4 & 30.1 $\pm$ 0.3 & \\
UDF12-3895-7114 & 3:32:38.95 & -27:47:11.4 & 8.6 $^{+0.8}_{-0.6}$ & $>$ 30.9 & 30.4 $\pm$ 0.5 & 30.1 $\pm$ 0.3 & 30.1 $\pm$ 0.4 & \\
UDF12-3947-8076 & 3:32:39.47 & -27:48:07.6 & 8.6 $^{+0.2}_{-0.2}$& 31.0 $\pm$ 0.5 & 29.5 $\pm$ 0.2 & 29.0 $\pm$ 0.1 & 29.0 $\pm$ 0.1 & UDFy-39468075 B11\tablenotemark{b}\\
\smallskip \\
\multicolumn{9}{c}{\bf Earlier Candidates\tablenotemark{a}} \smallskip \\
UDFj-39546284 & 3:32:39.54 & -27:46:28.4 & 11.9 $^{+0.3}_{-0.5}$ & $>$ 31.2 & $>$ 30.7 & $>$ 30.5 & 29.3 $\pm$ 0.2 & B11\tablenotemark{b} z$\simeq$10.3 \\
UDFj-38116243 & 3:32:38.11 & -27:46:24.3 & $-$ & $>$ 31.2 & $>$ 30.1 & 30.3 $\pm$ 0.5 & 30.0 $\pm$ 0.3 & B UDF09 \tablenotemark{c} \#1, B11\tablenotemark{b} \#2 \\
UDFj-43696407 & 3:32:43.69 & -27:46:40.7 & 7.6 $^{+0.4}_{-0.6}$ & 31.0 $\pm$ 0.6 & $>$ 30.1 & 29.9 $\pm$ 0.3 & 29.5 $\pm$ 0.2 & B UDF09 \tablenotemark{c} \#2 \\
UDFj-35427336 & 3:32:35.42 & -27:47:33.6 & 7.9 $^{+0.9}_{-0.8}$ & $>$ 30.8 & 30.3 $\pm$ 0.4 & 30.2 $\pm$ 0.4 & 29.6 $\pm$ 0.2 & B UDF09 \tablenotemark{c} \#3 \\
UDFy-38135539 & 3:32:38.13 & -27:45:53.9 & 8.3 $^{+0.2}_{-0.1}$ & 30.1 $\pm$ 0.2 & 28.6 $\pm$ 0.1 & 28.5 $\pm$ 0.1 & 28.4 $\pm$ 0.1 & B11\tablenotemark{b} 8.5$<z<$9.5 \\
UDFy-37796000 & 3:32:37.79 & -27:46:00.0 & 8.1 $^{+0.1}_{-0.2}$ & 29.8 $\pm$ 0.1 & 28.6 $\pm$ 0.1 & 28.7 $\pm$ 0.1 & 28.7 $\pm$ 0.1 & B11\tablenotemark{b} 8.5$<z<$9.5 \\
UDFy-33436598 & 3:32:33.43 & -27:46:59.8 & 7.9 $^{+0.2}_{-0.3}$ & 30.3 $\pm$ 0.4 & 29.3 $\pm$ 0.2 & 29.4 $\pm$ 0.2 & 29.4 $\pm$ 0.1 & B11\tablenotemark{b} 8.5$<z<$9.5\smallskip \\
\enddata
\tablenotetext{a}{Upper photometric limits are 2 $\sigma$.}
\tablenotetext{b}{Bouwens et al. (2011)}
\tablenotetext{c}{Bouwens et al. UDF09 yr. 1}
\end{deluxetable*}

\begin{figure*}
\begin{center}
\includegraphics[width=0.8\textwidth]{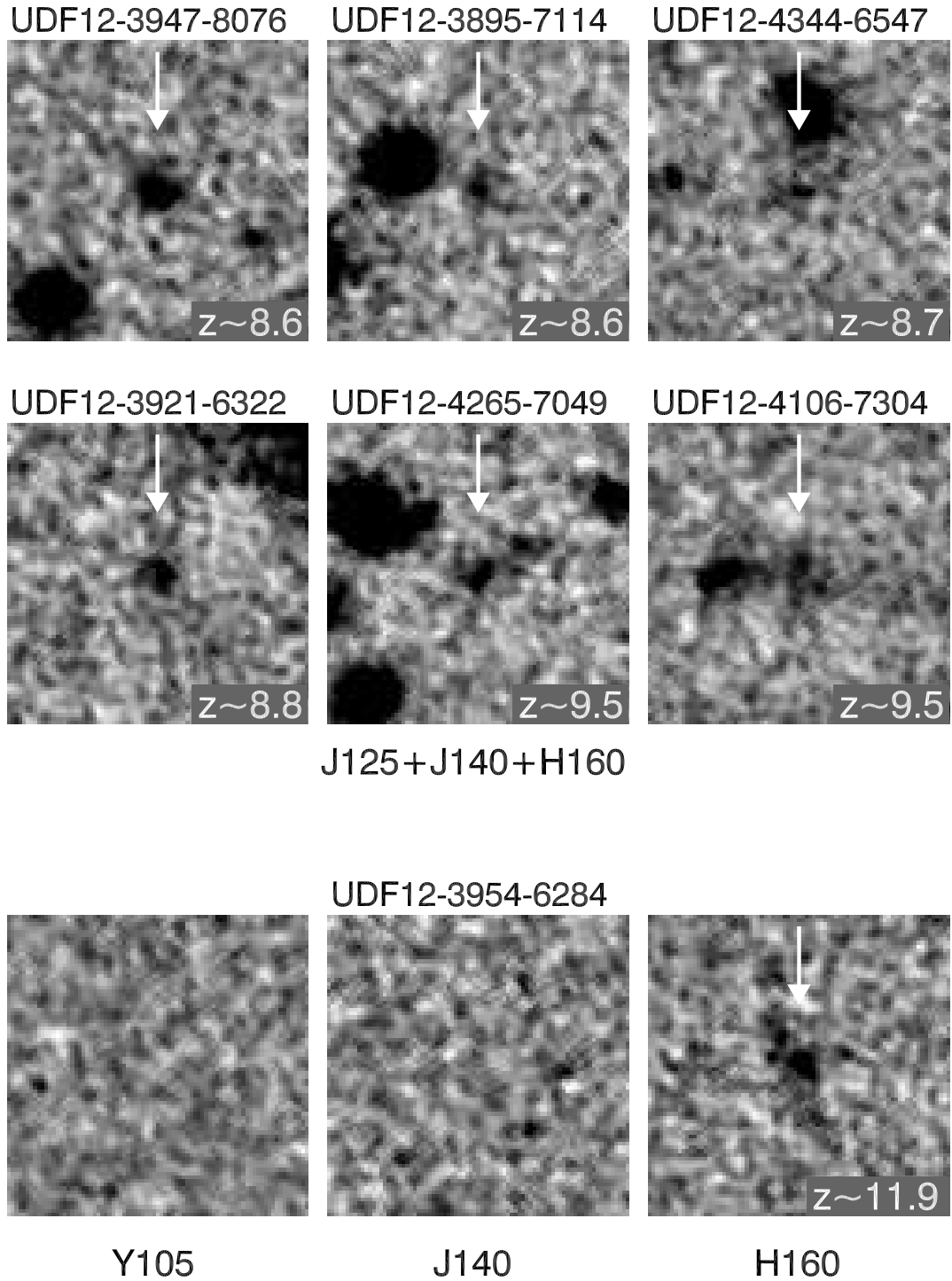}
\caption{Hubble Space Telescope WFC3/IR images of the promising $z>8.5$ candidates
from combined UDF12 and earlier data. Each panel is 2.4 arcsec on a side.
(Top two rows) Summed (F125W+F140W+F160W) images for 6 sources with $8.5<z<$10.0.
(Bottom row)  F105W, F140W, F160W images for UDF12-3954-6284$\equiv$UDFj-39546284\textbf{•}.
}
\end{center}
\end{figure*}

\begin{figure*}
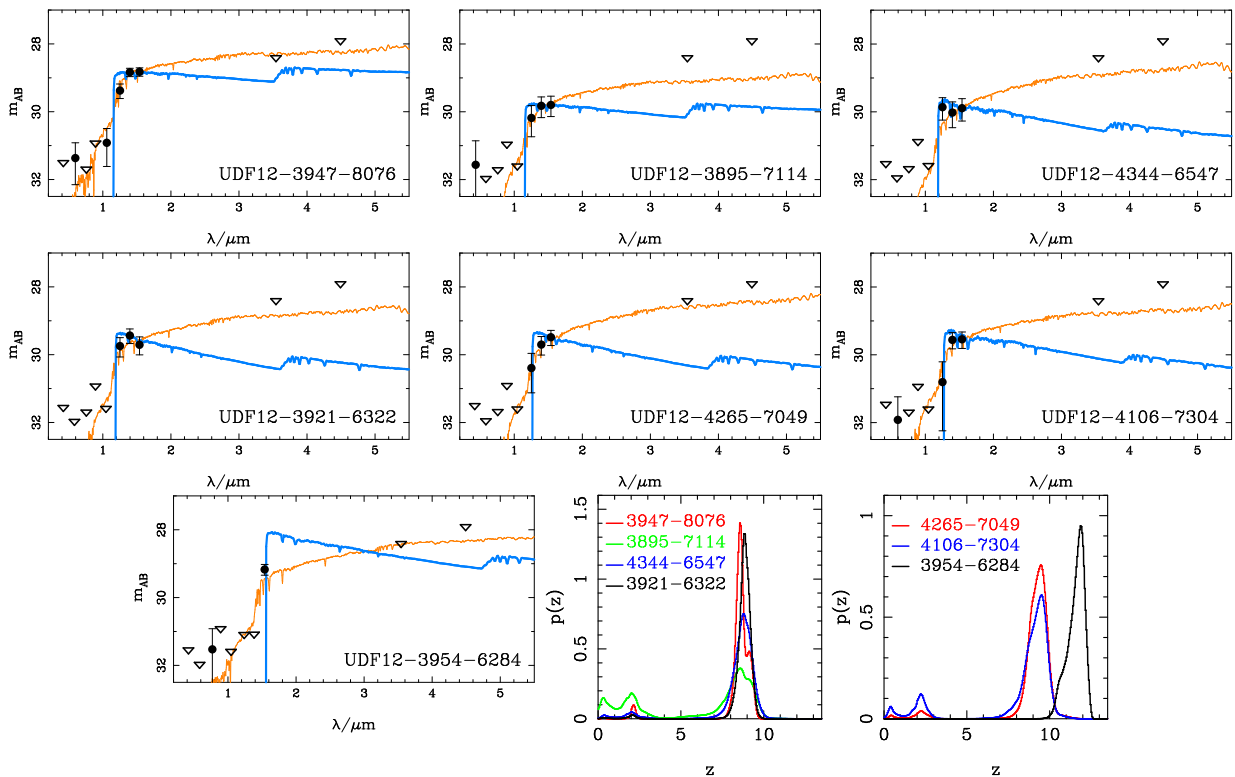

\begin{center}
%


\includegraphics[width=0.17\textwidth,angle=-90]{udf12-3947-8076_revised.ps}
\includegraphics[width=0.17\textwidth,angle=-90]{udf12-3895-7114_revised.ps}
\includegraphics[width=0.17\textwidth,angle=-90]{udf12-4344-6547_revised.ps}
\includegraphics[width=0.17\textwidth,angle=-90]{udf12-3921-6322_revised.ps}
\includegraphics[width=0.17\textwidth,angle=-90]{udf12-4265-7049_revised.ps}
\includegraphics[width=0.17\textwidth,angle=-90]{udf12-4106-7304_revised.ps}
\includegraphics[width=0.17\textwidth,angle=-90]{udf12-3954-6284_revised.ps}
\includegraphics[width=0.2\textwidth,angle=-90]{pz1_revised.ps}
\includegraphics[width=0.2\textwidth,angle=-90]{pz2_revised.ps}

\caption{\label{fig:seds}Spectral energy distributions for 7 promising
$z>$8.5 candidates from combined UDF12 and earlier data. Blue lines
represent the adopted high $z$ solution, orange lines the best rejected
low $z$ alternative. Upper limits are 1-$\sigma$. IRAC limits of $m_{3.6} > 28.5$
and $m_{4.5} > 28.0$ are based on a
deconfusion analysis of the \citet{labbe2012} data using the 
UDF12 $H_{160}$ image and the technique described in \citet{mclure2011a}. 
The final two panels show photometric likelihood fits for the sub-samples with $8.5<z<9.5$ 
and $z\geq9.5$.}
\end{center}
\end{figure*}

\begin{figure*}
\begin{center}
\includegraphics[width=0.4\textwidth]{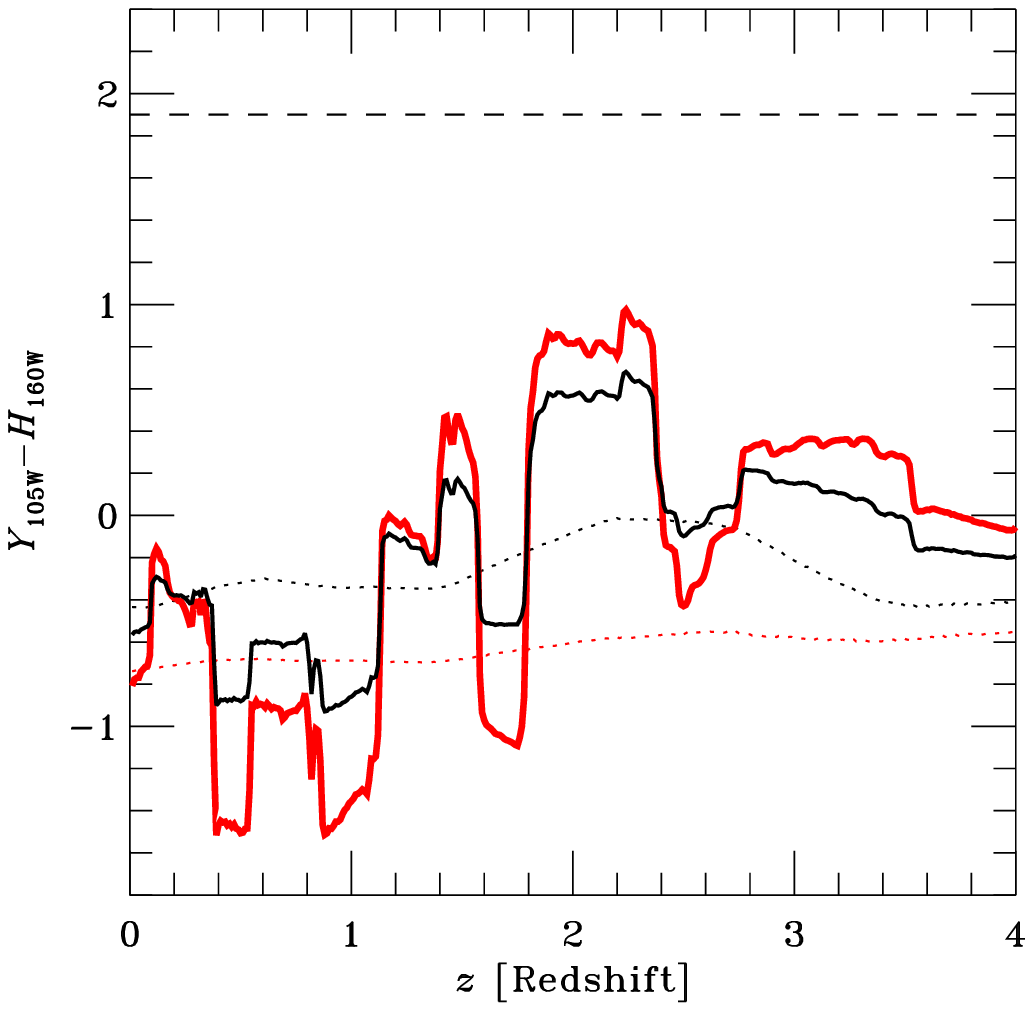}
\includegraphics[width=0.4\textwidth]{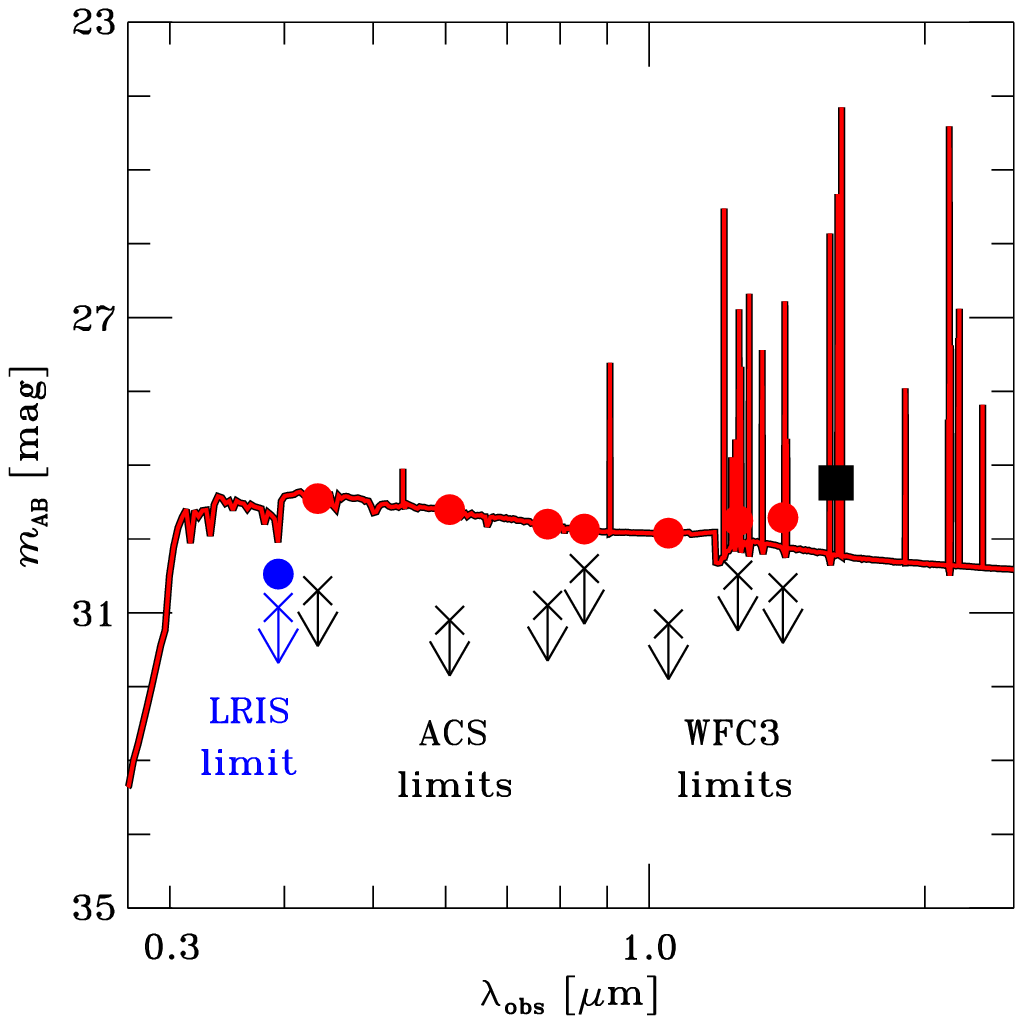}
\caption{Possible contamination by foreground extreme emission line galaxies.
(Left) F105W minus F160W color as a function of redshift for a 1 Myr (red) and 10 Myr (black) 
metal-poor dust-free stellar population incorporating nebular emission following the precepts
of Ono et al (2010). Expectations for a stellar continuum only are shown by
the dotted lines. The upper dashed line is the lower limit for UDFj-39546284 derived 
from the 2$\sigma$ F105W limit and the $>$6$\sigma$ F160W detection.
(Right). Simulated spectrum of UDFj-39546284 assuming a 10 Myr starburst at $z$=2.24 
where [O III] emission dominates the F160W signal. Arrows indicate 2$\sigma$ upper limits 
from non-detections in the various WFC3/IR and ACS  bands. The blue circle indicates the 
expected Lyman $\alpha$ strength rejected by a Keck spectrum (blue arrow).}
\end{center}
\end{figure*} 

\begin{figure*}
\begin{center}
\includegraphics[width=0.6\textwidth]{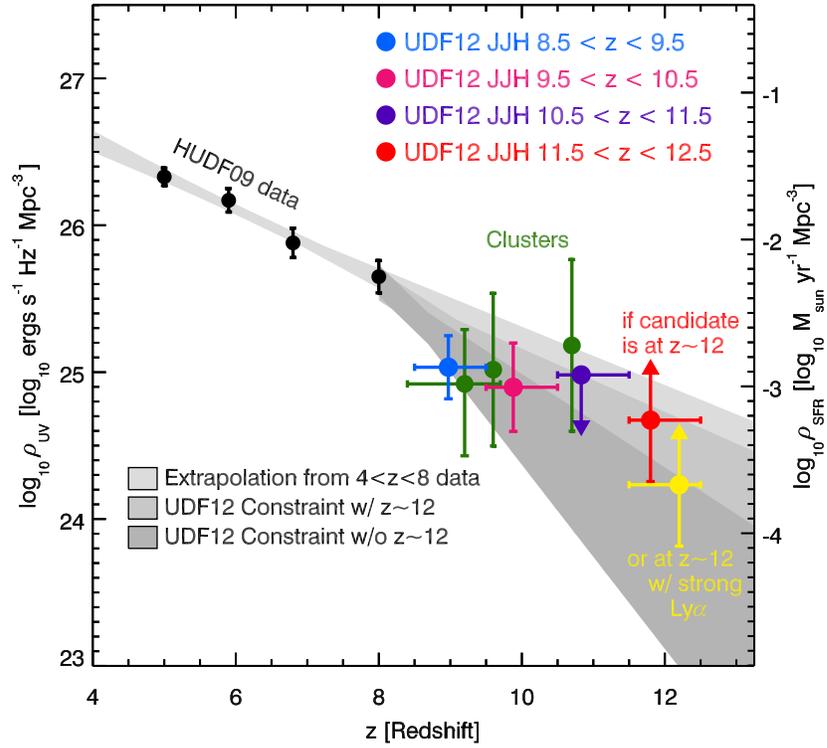}
\caption{\label{fig:rho_uv}Luminosity and star formation rate (SFR) density versus 
redshift inferred from UDF12. Reddening corrected luminosity densities are shown 
from Bouwens \etal\ (2007, 2011) over the redshift range 5$<z<$8 (black points).  Extrapolating their
evolution to redshift $z\sim13$ provides the lightest gray area.
Claimed estimates from the CLASH detections (green points)\citep{zheng2012a, coe2012a, bouwens2012b}
are shown. Luminosity densities are shown for the four $8.5\lesssim z\lesssim9.5$ sources 
(blue data point) and the two $9.5\lesssim z\lesssim10.5$ objects (magenta point).
The nondetection at $10.5\lesssim z \lesssim 11.5$ provides an upper limit at $z\approx10.8$ (purple limit).
The single $z\sim12$ source provides a conservative lower limit at $z\approx11.8$ (red point).  If this
source has strong Ly$\alpha$ emission, the luminosity density limit becomes the yellow point.
Overlapping maximum likelihood 68\% confidence regions on a linear trend in the luminosity density with 
redshift from $z\sim8$ are shown with (medium gray) and without (dark gray) the $z\sim12$ object.
The luminosity density computation is described in Section \ref{sec:abundance}. Associated star formation
rates (right axis) were calculated using the conversion of \citet{madau1998a}. }
\end{center}
\end{figure*}

\end{document}